\documentclass[pre, amsmath, amssymb, superscriptaddress]{revtex4}

\usepackage{multirow}
\usepackage{graphicx}
\usepackage{dcolumn}
\usepackage{bm}
\usepackage{braket}
\usepackage{xcolor}

\usepackage{amsmath}
\usepackage{SIunits} 

\newcommand{\beq}{\begin{equation}}
\newcommand{\eeq}{\end{equation}}

\usepackage{xr}
\usepackage{caption}

\begin{document}

\title{
Insights into the role of dynamical features in protein complex formation: the case of SARS-CoV-2 spike binding with ACE2
}

\author{Greta Grassmann  \footnote{\label{corr} For correspondence write to: greta.grassmann@uniroma1.it}}
\affiliation{Department of Physics, Sapienza University of Rome, Rome, Italy}
\affiliation{Center for Life Nano \& Neuro Science, Italian Institute of Technology, Rome, Italy}

\author{Mattia Miotto}
\affiliation{Center for Life Nano \& Neuro Science, Italian Institute of Technology, Rome, Italy}

\author{Francesca Alessandrini}
\affiliation{Department of Physics, Sapienza University of Rome, Rome, Italy}

\author{Leonardo Bo'}
\affiliation{Center for Life Nano \& Neuro Science, Italian Institute of Technology, Rome, Italy}

\author{Giancarlo Ruocco}
\affiliation{Center for Life Nano \& Neuro Science, Italian Institute of Technology, Rome, Italy}
\affiliation{Department of Physics, Sapienza University of Rome, Rome, Italy}

\author{Edoardo Milanetti}
\affiliation{Department of Physics, Sapienza University of Rome, Rome, Italy}
\affiliation{Center for Life Nano \& Neuro Science, Italian Institute of Technology, Rome, Italy}

\author{Andrea Giansanti}
\affiliation{Department of Physics, Sapienza University of Rome, Rome, Italy}
\affiliation{Istituto Nazionale di Fisica Nucleare, Roma, Italy}


\begin{abstract}
The functionality of protein–protein complexes is closely tied to the strength of their interactions, making the evaluation of binding affinity a central focus in structural biology. However, the molecular determinants underlying binding affinity are still not fully understood. In particular, the entropic contributions, especially those arising from conformational dynamics, remain poorly characterized.\\
In this study, we explore the relationship between protein motion and binding stability and its role in protein function.
To gain deeper insight into how protein complexes modulate their stability, we investigated a model system with a well-characterized and fast evolutionary history: a set of SARS-CoV-2 spike protein variants bound to the human ACE2 receptor, for which experimental binding affinity data are available.
Through Molecular Dynamics simulations, we analyzed both structural and dynamical differences between the unbound (apo) and bound (holo) forms of the spike protein across several variants of concern.\\
Our findings indicate that a more stable binding is associated with proteins that exhibit higher rigidity in their unbound state and display dynamical patterns similar to that observed after binding to ACE2. 
The increase of binding stability is not the sole driving force of SARS-CoV-2 evolution. More recent variants are characterized by a more dynamical behavior that determines a less efficient viral entry but could optimize other traits, such as antibody escape.\\
These results suggest that to fully understand the strength of the binding between two proteins, the stability of the two isolated partners should be investigated.

\end{abstract}


\flushbottom
\maketitle

\section*{Introduction}
Understanding the molecular mechanisms that regulate protein binding remains a fundamental yet unresolved challenge in computational biology \cite{grassmann2024computational, lu2020recent,grassmann2023electrostatic, grassmann2025compact,grassmann2025exploring}. Analyzing protein-protein interactions involves multiple aspects, including determining whether two proteins interact, identifying their binding sites, predicting their binding pose, and assessing the stability of the resulting complex. 
The strength related to the binding interaction is commonly quantified by the binding affinity ($B_a$).
Its evaluation is crucial because it is associated with protein complex functionality \cite{kamada2014prediction}. It has many applications, especially in applied fields like protein design~\cite{DiRienzo2021_, DeLauro2022}, where it helps evaluate how residue mutations affect molecular recognition and binding or rank the binding poses generated by protein–protein docking algorithms.
Binding affinity is commonly described with the experimental measure of the dissociation constant ($K_d$) \cite{jost2020quantifying,wang2022highly} through the relationship $B_a =-\log_{10}K_d$: a lower $K_d$ indicates stronger binding, while a higher $K_d$ suggests weaker interactions.\\ 
Despite decades of theoretical and computational advancements in predicting protein-protein binding affinity, as reviewed by Siebenmorgen \textit{et al.} \cite{siebenmorgen2020computational}, the problem remains highly complex and unsolved. The dissociation constant is directly related to the Gibbs free energy change upon binding \cite{kastritis2013binding}, according to the equation $\Delta G = RT \ln(K_d)$, where $R$ is the universal gas constant (8.3144 $\frac{J}{K mol}$) and $T$ the absolute temperature. The Gibbs free energy can also be expressed as $\Delta G = \Delta H - T\Delta S$, where $\Delta H$ and $\Delta S$ represent the changes in enthalpy and entropy of the system upon complex formation, respectively.
Enthalpy ($\Delta H$) accounts for the total energetic variation in the system and includes contributions from both the solute and solvent internal energies, as well as the energy required to establish the system’s physical organization. Upon binding, enthalpic changes result from (i) the formation of noncovalent interactions at the interface (e.g., van der Waals contacts, hydrogen bonds, ionic interactions, and hydrophobic effects), (ii) the loss of solute–solvent interactions, and (iii) the reorganization of the solvent molecules around the emerging complex.
Entropy ($\Delta S$), a measure of the system’s disorder or the number of accessible microstates, can be decomposed into distinct contributions: conformational entropy ($\Delta S_{conf}$), solvation entropy ($\Delta S_{sol}$), rotational–translational entropy ($\Delta S_{r-t}$), and other possible components ($\Delta S_{other}$), such as those arising from protonation changes.\\
The enthalpic contribution to binding has been well characterized thanks to detailed atomic-resolution structural models \cite{desantis2022spatial}, as well as the solvation entropic contribution \cite{dill1990dominant,hilser2006statistical}.
Other entropic components, particularly conformational entropy, remain more elusive. This is due to the intrinsic complexity of protein motions, which involve a rugged energy landscape with numerous local minima, anharmonic fluctuations, and complex couplings between internal degrees of freedom.
Nevertheless, studies based on NMR-derived methyl order parameters have provided evidence that conformational entropy can be comparable in magnitude to solvation entropy \cite{lipari1982model}. In a dataset of 28 protein–protein complexes, NMR relaxation methods, used as a proxy for backbone and side-chain flexibility, revealed that $\Delta S_{conf}$ significantly contributed to high-affinity binding in approximately 25\% of cases \cite{caro2017entropy}. However, these experimental approaches do not directly measure correlated dynamics, protein main chain motions, or angular dynamics, but rather reconstruct them from empirical, linear fits across sets of reference molecules.\\
This highlights the importance of computational strategies that can directly capture protein dynamics with atomic resolution~\cite{Miotto2020_}, such as MD simulations.
MD simulations are the basis of several widely used approaches for estimating binding affinities, including MM/PB(GB)SA methods \cite{valdes2021gmx_mmpbsa}, alchemical free energy calculations \cite{song2020evolution}, and more recently, deep learning–based techniques \cite{min2024static}. These approaches analyze the conformational ensemble sampled during the MD trajectory of the complex, capturing entropic contributions that are critical to binding affinity.
However, they typically overlook the dynamic pathways leading to complex formation.\\
In this context, we present a comprehensive computational study based on Molecular Dynamics (MD) simulations to further explore the link between protein motion and binding affinity. 
We investigated as a case study the interaction between the human Angiotensin-converting enzyme 2 (ACE2) receptor and the Severe Acute Respiratory Syndrome Coronavirus 2 (SARS-CoV-2). Specifically, we examine the SARS-CoV-2 spike (S) glycoprotein, which mediates viral attachment and entry into host cells via its Receptor Binding Domain (RBD) \cite{yan2020structural,cerutti2021potent}.
This system is of significant scientific interest due to the onset of Coronavirus Disease 2019 (COVID-19) \cite{zhou2020pneumonia,abdool2021new} and serves as a valuable model for exploring the dynamic and structural properties that influence protein binding. Like other coronaviruses, SARS-CoV-2 undergoes random mutations during viral replication, with those conferring increased fitness being selectively retained, leading to the emergence of new variants \cite{domingo1997rna, PhysRevResearch.2.043026, duchene2020temporal, portelli2020exploring, Miotto_2022}. During the COVID-19 pandemic, several variants of concern have been identified, each with distinct mutations in the S protein. Each variant can be compared with the wild-type (WT) form of the S protein, and the observed differences can be related to changes in binding affinity that have been experimentally determined.
The deep investigations performed on the evolutionary dynamics of the S-ACE2 complex make it a perfect case of study for the investigation of how dynamical features affect protein binding.\\
For this study, we selected the alpha, beta, gamma, delta, and omicron variants. Their experimentally measured $K_d$ were reported by Han \textit{et al.} \cite{han2022receptor}, alongside that of the WT S protein. The values used in this study are listed in Table \ref{tab_kd} of the Methods.
Interestingly, binding affinity does not consistently increase over the course of SARS-CoV-2 evolution. This is because viral fitness is not solely driven by stronger ACE2 binding \cite{miotto2023differences,miotto2022inferring}. Instead, multiple factors (such as immune evasion, transmission efficiency, and viral replication) collectively shape the evolutionary trajectory of the virus \cite{junker2022covid}. The selection pressure acting on SARS-CoV-2 appears to favor variants that optimize overall viral efficiency rather than simply maximizing binding affinity.\\
Among the variants analyzed, the first to emerge was alpha (lineage B.1.1.7), detected in November 2020 \cite{faria2021genomic}. This variant carries the N501Y mutation in its RBD \cite{harrington2021confirmed} and was estimated to be 40–80\% more transmissible than the WT \cite{lin2021disease,kow2021mortality}. Alpha has one of the lowest $K_d$ values (5.40 nM), indicating a strong interaction with ACE2.
One month later, the beta variant (B.1.351) was identified. Beta exhibits a higher $K_d$ (13.83 nM) and an enhanced ability to bind human cells, likely due to three key RBD mutations (N501Y, K417N, E484K) \cite{grabowski2021sars}.
The delta variant (B.1.617.2), detected in late 2020, harbors L452R and T478K mutations in the RBD~\cite{adam2021scientists}. It spread faster than both the WT and alpha variant \cite{campbell2021increased}, becoming the dominant strain by June 2021, according to the World Health Organization (WHO). Delta has a relatively high $K_d$ of 26.07 nM.
The gamma variant, first identified in January 2021, contains 17 amino acid substitutions, including three in the RBD (N501Y, E484K, K417T) \cite{faria2021genomic}, and has a $K_d$ of 11.00 nM.
The latest variant considered in this study, omicron (B.1.1.529), was detected in November 2021 \cite{gowrisankar2022omicron}. Omicron exhibited a higher reinfection risk than previous strains \cite{vitiello2022advances} and introduced numerous novel mutations. Compared to the WT S protein, omicron harbors 30 amino acid substitutions, three deletions, and one insertion in the S protein, with 15 mutations located in the RBD \cite{vitiello2022advances}. It has the highest $K_d$ among the analyzed variants (31.40 nM).\\

We performed MD simulations of the S protein in both its isolated (apo) and ACE2-bound (holo) forms for each variant and the WT. By comparing the structural and dynamical properties of the apo and holo conformations, we observed that greater stability in the unbound form often correlates with higher stability in the bound form. This increased stability is accompanied by a more consistent conformational exploration across apo and holo simulations. On the other hand, S-ACE2 complexes with higher $B_a$ values (stronger binding) tend to have a more variable intermolecular distance.\\
These findings indicate that the S protein can adopt different binding mechanisms. 
Variants with lower binding affinity tend to follow a binding mode resembling the induced-fit model, where binding is mediated by conformational changes \cite{csermely2010induced}. Although this mode leads to weaker binding, the resulting structural flexibility is likely to confer advantages in other aspects of viral fitness, such as enhanced antibody escape \cite{di2023dynamical}.
Conversely, variants that prioritize efficient viral entry into host cells appear to shift toward a more rigid, lock-and-key-like mechanism, characterized by minimal structural rearrangements upon receptor binding \cite{csermely2010induced}.

\section*{Results}
As said above, to investigate the role of conformational entropy in modulating binding strength, we focused on a system whose evolutionary trajectory is well characterized in terms of mutation-induced effects on protein structure and receptor affinity: the complex formed by S protein variants and the ACE2 cellular receptor.
Using MD simulations, we analyzed five selected variants along with the WT, both in their unbound and ACE2-bound forms. This approach allowed us to systematically compare changes in flexibility and stability between the two states across variants exhibiting a range of binding affinities.

\subsection*{Flexibility changes upon binding}
\begin{figure*}[]
\centering
\includegraphics[width=\linewidth]{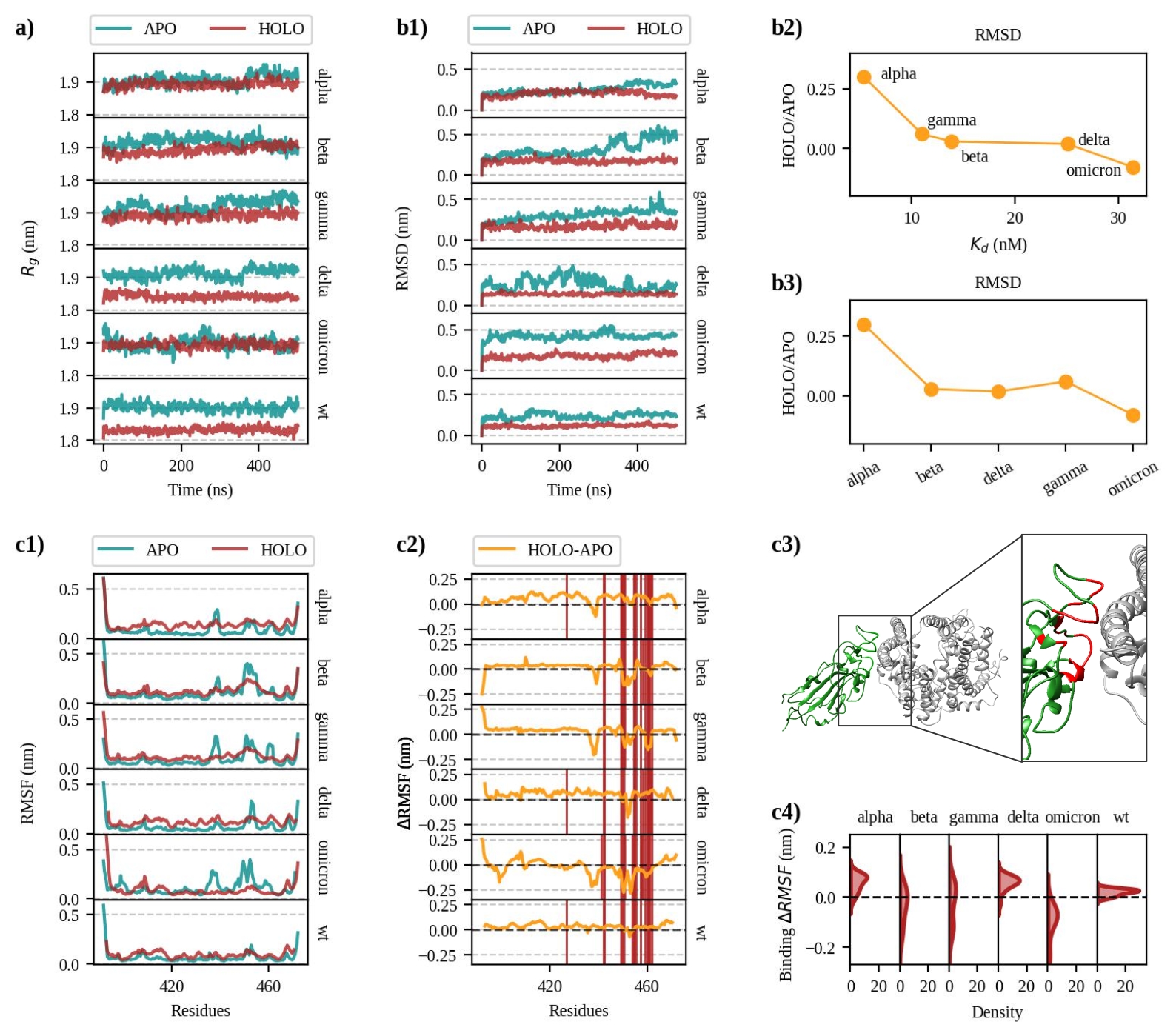}
\caption{\textbf{Comparison of the motion of apo and holo structures.}
\textbf{a)} Radius of gyratio ($R_g$) as a function of time for the apo (blue) and holo (red) structures.
\textbf{b1)} For each variant and the WT, RMSD as a function of time for the apo (green) and holo (red) structures. The RMSD is computed against the starting minimized structure, considering the proteins backbone.
\textbf{b2)} Difference in the mean RMSD values in b1) between holo and apo structures for each variant, plotted as a function of the dissociation constant $K_d$ measured for that variant in complex with ACE2. The difference in mean RMSD for WT was subtracted from each variant’s value. Variants are ordered by increasing $K_d$.
\textbf{b3)} The values shown in b2 are ordered according to the chronological order of observation.
\textbf{c1)} Root Mean Square Fluctuation (RMSF) per residue for the apo (green) and holo (red) structures.
\textbf{c2)} Difference in RMSF ($\Delta$RMSF) per residue between holo and apo structures, based on the values in c1. Red vertical lines indicate the interacting residues for each variant and WT.
\textbf{c3)} For each variant and the WT, the S interacting residues are selected based on the frame where the RMSD is closest to the mean RMSD observed during that complex simulation. Interacting residues are defined as those with centroids within 8\AA~from the other chain residues' centroids. The graph shows an example: the alpha S (green) in complex with ACE2 (grey). The insert highlight S residues interacting with ACE2 in red.
\textbf{c4)} Distributions of $\Delta$RMSF from c2) for the interacting residues highlighted in red in c3.
}
\label{fig1}
\end{figure*}

To compare the motion of apo and holo conformations across the five variants and the WT, we analyzed the variation in the radius of gyration ($R_g$) throughout the simulations. As shown in Figure \ref{fig1}a, the mean $R_g$ of all apo conformations is approximately 1.9 nm. A similar trend is observed for the holo conformations of the alpha, beta, gamma, and omicron variants. However, the delta and WT holo conformations exhibit a lower $R_g$ compared to their respective apo states, with a mean value of approximately 1.83 nm. Since $R_g$ is an indicator of protein compactness, this suggests that delta and WT conformations become more folded upon binding.\\
Figure \ref{fig1}b1 illustrates that all holo conformations display a comparable, steady behavior in terms of Root Mean Square Deviation (RMSD), with mean values of 0.2$\pm$0.03 nm, 0.17$\pm$0.02 nm, 0.17$\pm$0.03 nm, 0.13$\pm$0.01 nm, 0.18$\pm$0.03 nm, and 0.12$\pm$0.01 nm for alpha, beta, gamma, delta, omicron, and WT, respectively. In contrast, apo conformations exhibit greater RMSD fluctuations: the standard deviation of the apo trajectory is four times higher than the holo counterpart for beta, five times higher for delta, and twice as high for WT.
As expected, ACE2 binding stabilizes the S protein, as evidenced by lower RMSD values in holo trajectories compared to apo ones. This stabilization effect intensifies with increasing dissociation constant ($K_d$) values, as depicted in Figure \ref{fig1}b2.
Here, the ratio of the mean RMSD values between holo and apo trajectories (after subtracting the WT value from each variant) is plotted against the dissociation constant $K_d$ of each variant, as listed in Table \ref{tab_kd}.
Since a higher $K_d$ corresponds to weaker binding affinity, this trend suggests that the stabilizing effect of binding diminishes as binding affinity increases: the stability of the apo conformation appears to have been optimized over the course of SARS-CoV-2 evolution toward increased binding affinity for its cellular receptor. This evolutionary trend follows the chronological emergence of variants (see Figure \ref{fig1}b3).\\
To further assess the stability differences between the apo and holo conformations, we computed the S protein residues Root Mean Square Fluctuation (RMSF). The RMSF of a residue represents the average deviation of its atomic positions from their mean positions over time, thereby providing a residue-level insight into local structural flexibility. Figure \ref{fig1}c1 displays the RMSF values for each S residue in both conformational states, as derived from MD simulation at equilibrium. 
Notably, the apo trajectories exhibit pronounced peaks in RMSF for specific residues, indicating localized flexibility, whereas the holo conformations show a more uniform fluctuation profile.
This trend is further illustrated in Figure \ref{fig1}c2, which reports the per-residue differences in RMSF between holo and apo conformations ($\Delta$RMSF), highlighting a reduction in flexibility for specific residues upon binding, as indicated by negative $\Delta$RMSF peaks.
This binding-induced stabilization is particularly evident in the highlighted regions comprising ACE2-interacting residues (as defined in the Methods and exemplified for the alpha variant in Figure \ref{fig1}c3).
Figure \ref{fig1}c4 shows the distribution of $\Delta$RMSF values restricted to these residues. For the beta, gamma, and omicron variants, binding leads to a clear reduction in the mobility of the interface, with omicron showing the strongest effect (mean $\Delta$RMSF = –0.1 nm).
In contrast, a positive mean $\Delta$RMSF is observed for alpha, delta, and WT (0.05 nm, 0.057 nm, and 0.02 nm, respectively), indicating slightly increased mobility after binding.
Nonetheless, in all cases, the mean $\Delta$RMSF of the interacting residues is lower than the average observed across all residues, suggesting that ACE2 binding specifically stabilizes the interface more than the rest of the S protein.

\subsection*{Covariant motion analysis as a measure of the variation in holo and apo conformations}

\begin{figure*}[]
\centering
\includegraphics[width=\linewidth]{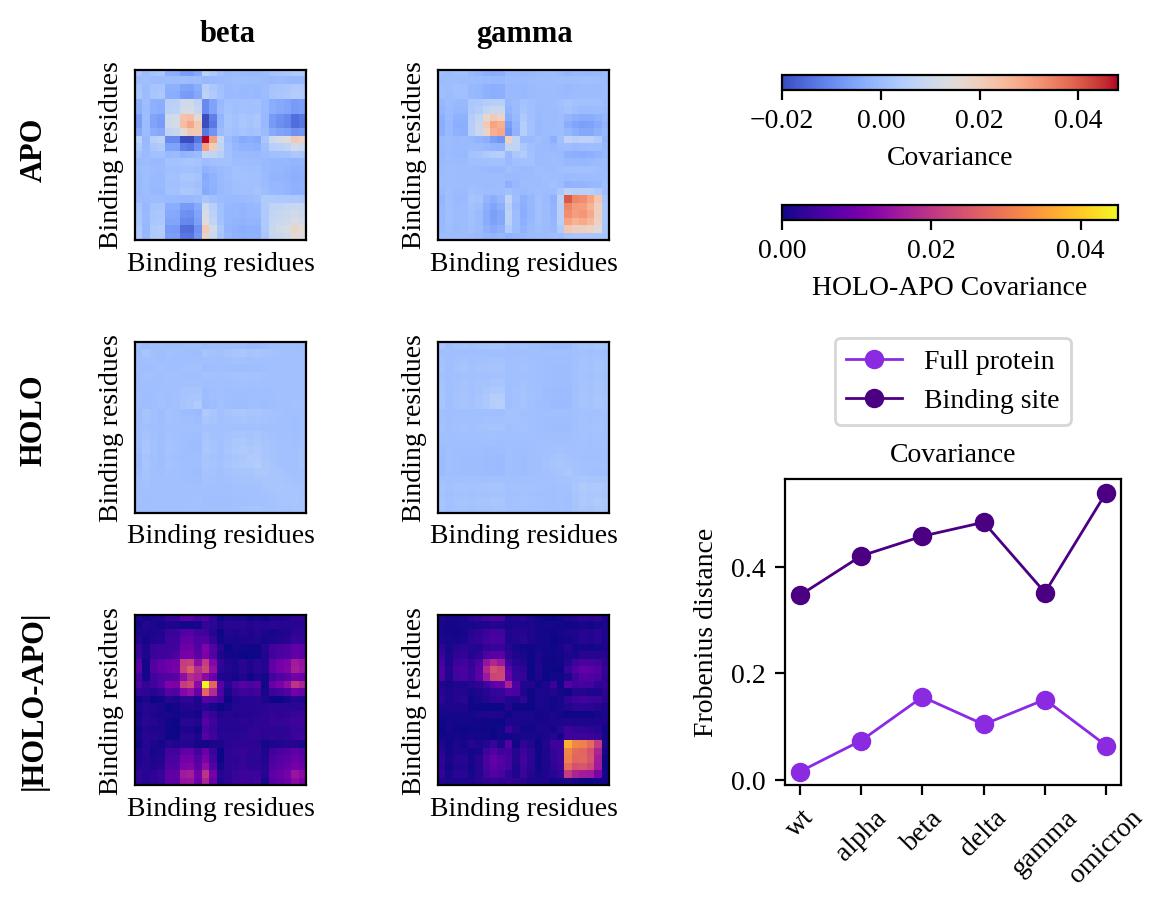}
\caption{\textbf{Covariant analysis of S binding site residues motion in apo and holo conformations.}
The covariance matrices of the beta and gamma S residues interacting with ACE2 (listed in Table \ref{tab_big_patch}) are presented as examples in the first two columns. The top row displays covariance values computed from the apo conformation simulation, while the middle row corresponds to the holo conformation. The colormap represents the covariance values, as indicated by the top colorbar.
The last row illustrates, for each residue pair, the difference in covariance between the holo and apo conformations. The bottom colorbar indicates the corresponding colormap values.
The same analysis was performed considering all residues and can be found in the Supplementary Information.
The plot in the bottom right compares the norm of the difference between holo and apo covariance matrices across all residues (light violet) and only interacting residues (dark violet) for the five variants and WT. These differences are quantified using the Frobenius distance and are ordered according to their chronological appearance.
}
\label{fig2}
\end{figure*}

To analyze the coordinated motion of S residues across all variants and the WT, we computed the covariance between atomic positions for all residue pairs in both holo and apo simulations. This resulted in a covariance matrix for each apo and holo structure.
Figure \ref{fig2} presents, as an example, the covariance matrices of the WT and omicron binding site residues (listed in Table \ref{tab_big_patch}), along with the norm of the differences between holo and apo conformations. To quantify these differences in dynamics, we computed the Frobenius distance between the holo and apo covariance matrices, as described in the Methods. Lower Frobenius distances indicate more similar matrices and, consequently, more comparable protein motions, where corresponding residue pairs exhibit similar covariance values.
The same Figure reports the Frobenius distances between holo and apo conformations for all variants and the WT. Covariances were calculated either for the residues forming the ACE2 binding site or for the whole chain. All covariance matrices are provided in the Supplementary Information.
As shown in Figure \ref{fig2}, the most pronounced differences in correlated motion occur at the binding site. The holo and apo covariance matrices of interacting residues exhibit a significantly higher Frobenius distance compared to those of the full S protein. This difference arises from the stronger covariance observed in the apo conformations: in all variants except delta, the maximum covariance in the apo state exceeds that of the holo state, with the most extreme case being gamma, where the highest apo covariance is more than seven times that of the holo trajectory.
This is in line with the reduction of the mean RMSD previously observed: the reduced motion of the residues determine a reduced correlated motion.
Also considering the reduction in the RMSF of interacting residues in holo conformations, as discussed in the previous section, these findings suggest that binding to ACE2 stabilizes and immobilizes the interface. This effect appears to become more pronounced over time across different variants, as illustrated in the bottom-right plot of Figure \ref{fig2}.

\subsection*{Principal component analysis of the covariance matrices highlights different behaviors of holo and apo conformations as a function of different binding affinity values}

\begin{figure*}[]
\centering
\includegraphics[width=\linewidth]{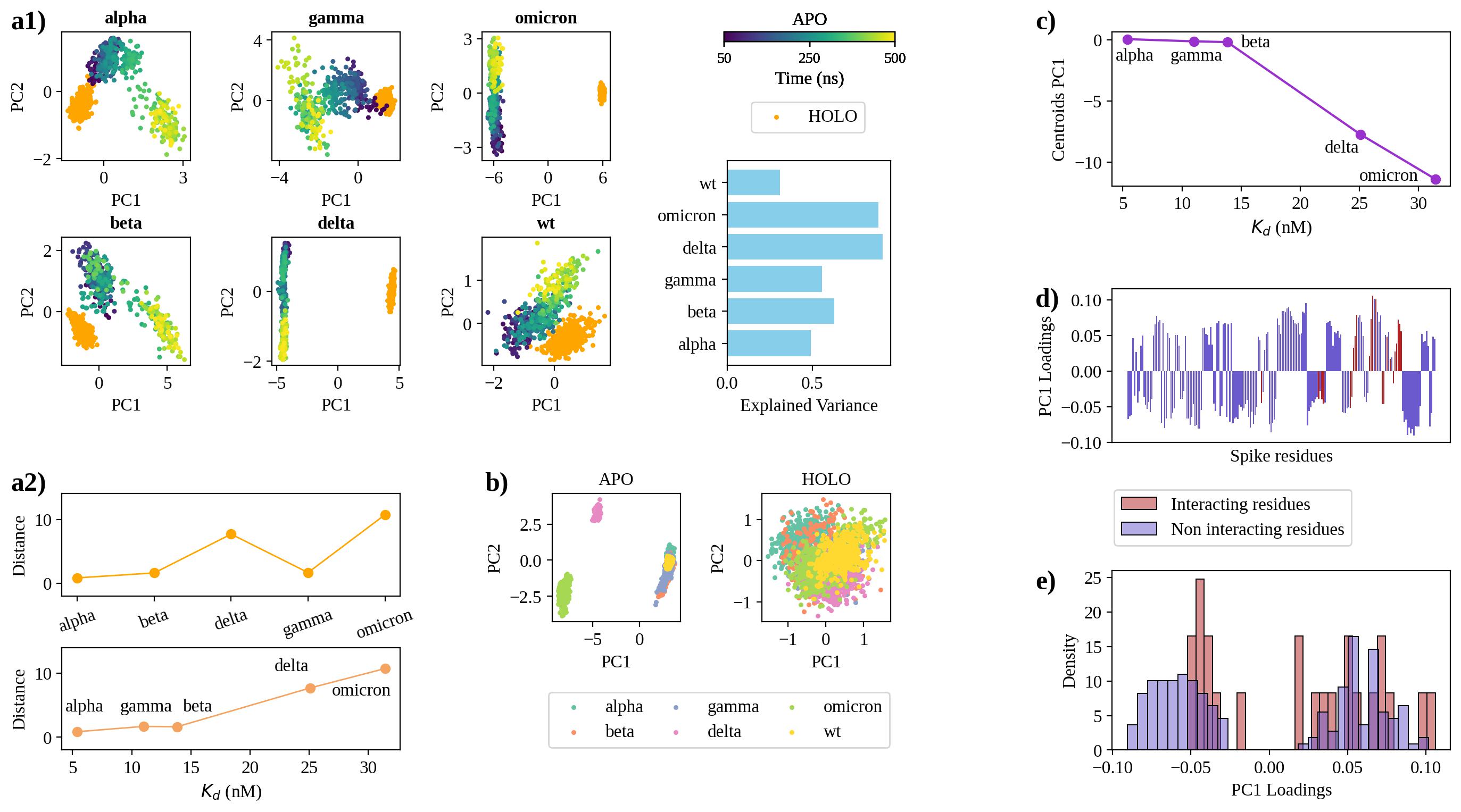}
\caption{\textbf{Principal Component Analysis of apo and holo trajectories.}
\textbf{a1)} Two-dimensional projection of the sampled conformations in the subspace spanned by the first two PCs of atomic positions covariances during the apo and holo simulation for each variant and the WT. 
Holo simulations points are colored in orange, while those from apo trajectories are colored from blue to yellow as simulation time goes from zero to 500 ns, as indicated by the colorbar. The first 50 ns of simulation were excluded from the analysis to account for the initial non-equilibrium phase of the system.
The bottom right plot shows the cumulative Explained Variance (EV) of the first two PCs for each case.
\textbf{a2)} For each plot in a1), the distance between the centroids of the apo and holo projections is computed.
The WT distance was subtracted from each variant’s value. 
The variants are ordered based on their time of observation (top) and their dissociation constant ($K_d$).
\textbf{b)} Two-dimensional projection of the sampled conformations in the subspace spanned by the first two PCs of the atomic positions covariances, comparing apo (left) and holo (right) simulation across all variants and WT. Each point is colored according to the corresponding variant, as indicated in the legend. 
\textbf{c)} For each variant, the centroid of the apo trajectory projection in the PC space shown in the left panel of b) is computed. The PC1 coordinate of each centroid is plotted as a function of the variant’s dissociation constant ($K_d$), with variants ordered by increasing $K_d$. The WT coordinate was subtracted from each variant’s value. 
\textbf{d)} The contribution (loading) of each S residue to the PC1 computed for the essential representation of the apo trajectories shown in the left panel of b is here reported. Red bars indicate interacting residues.
\textbf{e)} Histogram of the values shown in d for interacting (red) and non interacting (blue) residues.
}
\label{fig3}
\end{figure*}

We then analyzed the differences in dynamics between apo and holo simulations. For each variant and the WT, we projected the sampled configurations from the apo and holo simulations onto the essential space defined by the two PCs of the covariance matrix obtained from both trajectories, as shown in Figure \ref{fig3}a1.
Consistent with the stabilizing effect of binding observed in the RMSF and covariance analyses, holo conformations explore a more restricted conformational space compared to their apo counterparts. Moreover, the apo and holo structures do not tend to overlap.
The distance between the centroids of the two distributions is reported in Figure \ref{fig3}a2, plotted as a function of both the temporal progression of the variants (top) and their $K_d$ (bottom).
The WT distance was subtracted from each variant’s value. 
While the correlation with chronological appearance is less evident, variants with higher $K_d$ values than the WT (delta and omicron) show significantly greater centroid distances between their apo and holo projections. They also share a similar trend in essential space exploration, suggesting that S proteins undergoing more pronounced conformational changes upon binding tend to have lower binding affinities.\\
To further investigate the motion differences between variants and the WT in their holo and apo conformations, we performed a second PCA. This time, we defined a common essential space for all apo and holo structures and projected the configurations together, as shown in Figure \ref{fig3}b. Holo conformations cluster in overlapping regions, as expected, since they are all bound to the same ACE2 binding site. In contrast, apo conformations are more distinct, particularly in the case of delta and omicron, which occupy separate regions from each other and from the other variants and the WT. 
The same behavior is observed when projecting all structures onto a single essential space defined by two PCs that describe both apo and holo trajectories across all variants and the WT, as well as when only the interacting residues are considered (see Supplementary Information).\\
Interestingly, the distribution of variants in the essential space of the apo trajectories, as shown in Figure \ref{fig3}b, reflects their dissociation constant. By computing the centroid of each projection, we observe a shift in the PC1 coordinate of the centroid towards lower values as $K_d$ increases.
Figure \ref{fig3}c illustrates this trend by showing the difference between the PC1 centroid projection of each variant and that of WT, with variants ordered by increasing $K_d$. This observation suggests that variations in binding affinity are mirrored by changes in the essential motions of the S protein.\\
To further investigate the role of individual residues in these changes, we computed the loadings (i.e., the coefficients of the linear combination of the original variables defining the PCs) for each amino acid contributing to PC1 in the apo trajectories. The result is shown in Figure \ref{fig3}d, where the loadings obtained for the S residues interacting with ACE2 are highlighted in red.
As confirmed in Figure \ref{fig3}e, which shows a histogram comparing the loadings of interacting and non-interacting residues, the contribution of the binding site is not significantly higher than that of the rest of the protein. This is evidenced by the fact that the distribution of loadings for interacting residues does not shift toward higher absolute values.

\subsection*{Analysis of the spike-ACE2 complex shows the anomalous motion of the delta variant}

\begin{figure*}[]
\centering
\includegraphics[width=\linewidth]{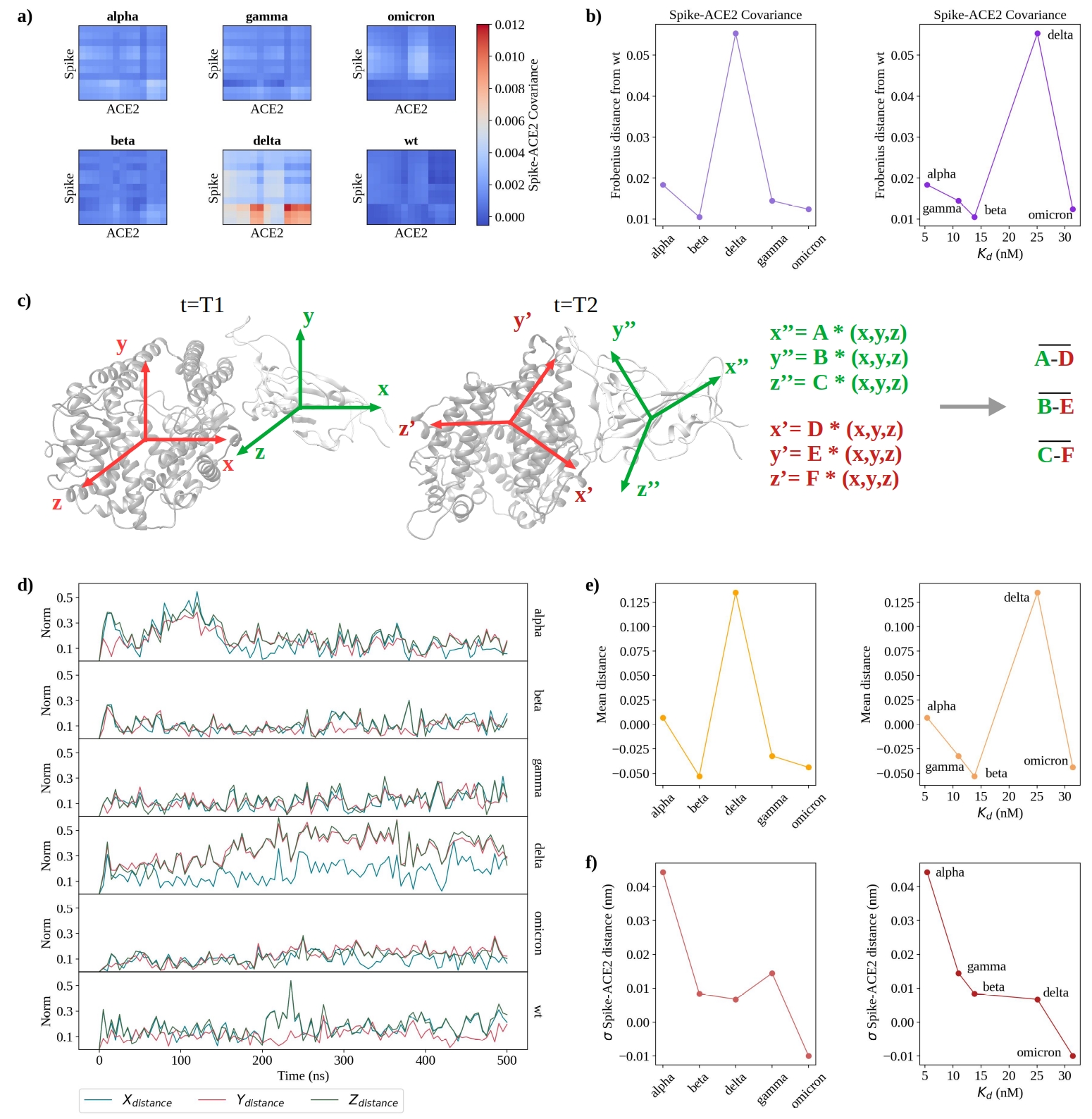}
\caption{\textbf{Comparison of spike and ACE2 motions.}
\textbf{a)} For each variant and the WT, the S and ACE2 sites that include residues identified as interacting in all cases are selected. The covariance matrix is then computed between these sites, with the colorbar indicating covariance values for all residues pairs.
\textbf{b)} The Frobenius distance quantifies the difference between the covariance matrices of each variant and the WT in a.
The left plot shows the Frobenius distance of each variant in order of their chronological appearance. The right plot orders variants according to their dissociation constant ($K_d$) for the S-ACE2 complex.
\textbf{c)} To evaluate the coordination of S and ACE2 rotation between two simulation frames (T1 and T2), rotation matrices are computed to fit the conformation of each protein from T2 to T1 using least squares.
This results, for each protein, in three vectors defining the new directions of the T2 coordinate system relative to T1 ($x,y,z$). In the figure, A, B, and C describe S rotation, while D, E, and F refer to ACE2 rotation. The difference between corresponding vectors pairs ($A-D$, $B-E$, $C-F$) measures the degree of coordination in their rotation.
\textbf{d)} For each variant and WT, the norm of the difference between the new directions of the $x$ (blue), $y$ (pink), and $z$ (green) axes ($\overline{{A-D}}$, $\overline{B-E}$, and $\overline{C-F}$ respectively in c) is plotted as a function of time.
\textbf{e)} For each complex in d, the average across simulation time and all three directions is computed. 
The WT average was subtracted from
each variant’s value.
The left plot orders the variants by their chronological appearance.
The right plot arranges them according to their $K_d$.
\textbf{f)} To analyze the relative motion of S and ACE2 during simulations, the centroids of the two chains are computed in all frames. 
The standard deviation ($\sigma$) of centroid distances is reported for each variant, having subtracted the value computed for the WT.
The left plot orders these values chronologically by variant appearance.
The right plot displays them as a function of the S-ACE2 complex $K_d$.
}
\label{fig4}
\end{figure*}

To characterize the differences between the variants and the WT, we analyzed the motion of the entire S-ACE2 complexes.
As a first step, we analyzed the covariance of residues located at the S–ACE2 binding interface, a region known for its high sequence conservation across SARS-CoV-2 variants \cite{milanetti2021silico}.
The specific residues included in this analysis were selected as detailed in the Methods section and are listed in Table \ref{tab_interfaces}.
Figure \ref{fig4}a presents the covariance matrices for the five variants and the WT. To quantify the differences between the variants and the WT, we computed the Frobenius distance between each variant’s covariance matrix and that of the WT. As shown in Figure \ref{fig4}b, most variants exhibit similar and relatively low Frobenius distances (lower than 0.02 nm), indicating that their S-ACE2 correlated motion is similar to that of the WT. However, delta stands out as an exception, with a Frobenius distance 3–5 times higher than those of the other variants, suggesting a distinct interaction pattern.\\
To further investigate this anomaly, we analyzed the relative motion of the S protein with respect to ACE2. Specifically, we evaluated the rotation of each chain during the simulation by computing, at each frame, the three vectors (denoted as A, B, and C for S protein, and D, E, and F for ACE2 in Figure \ref{fig4}c) that describe the new orientations of the x, y, and z axes relative to the initial conformation. 
The calculation was carried out using the $gmx~rotmat$ command in gromacs~\cite{gromacs}. The norm of the difference between the vectors describing the same axis in S and ACE2 provides a measure of how much the two proteins rotate together on that axis. Higher values indicate less coordinated rotation.
Figure \ref{fig4}d displays, for each variant and the WT, the norm of the vectors describing rotation around the x, y, and z axes. Delta exhibits the least coordinated motion between S and ACE2, with the highest values of vector norms. This trend is further summarized in Figure \ref{fig4}e, where the mean vector norms for each variant and the WT are reported. Delta has an average value of 0.3, whereas all other cases remain below 0.175.
Interestingly, the presence of stronger or weaker correlated motion does not show a clear dependence on the temporal emergence of the variants or their $K_d$. However, we observe that the variance of the centroid distance between S and ACE2 during the simulations decreases with increasing $K_d$ values. Figure \ref{fig4}f shows that while the variation in chain distance does not correlate with the temporal evolution of the variants, complexes with higher binding affinity (i.e., lower $K_d$) exhibit greater fluctuations in distance. See the Supplementary Information for a detailed analysis of centroid distances as a function of simulation time.

\section*{Conclusions}
To gain deeper insight into the mechanisms governing stable protein–protein interactions, we examined the SARS-CoV-2 S protein–ACE2 receptor complex as a case study, focusing on how this interaction has evolved over the course of the pandemic.
We analyzed structural and dynamical differences between the unbound (apo) and ACE2-bound (holo) forms of the S protein across multiple viral variants. In particular, we investigated how these differences correlate with changes in binding affinity.\\
As expected, the interaction with ACE2 stabilized both the WT and variant S proteins. However, the difference in global stability between the apo and holo conformations was less pronounced in variants with higher binding affinity. At a local level, increased structural variability was frequently observed in the apo state. RMSF analysis of the apo trajectories revealed multiple peaks, indicating regions with higher residue mobility, whereas the holo distributions were more uniform. Interestingly, in some S-ACE2 complexes (alpha, delta, and WT), the residues at the binding interface exhibited increased flexibility after binding. Nevertheless, in all cases, interacting residues in the holo conformations were consistently less mobile than the rest of the S protein.
Higher residue mobility in the apo state was also accompanied by greater covariance among residues, suggesting a more coordinated motion before binding. Upon ACE2 binding, this correlation decreased, particularly at the interface of the S protein. The reduction of correlated motion tended to be more pronounced in the later SARS-CoV-2 variants.\\
In the unbound state, all S proteins explored a broader region of conformational space, as demonstrated by a PCA capturing the essential motions of the protein chains. Conversely, when bound to ACE2, both the WT and variant S proteins exhibited more restricted conformational dynamics and converged toward more similar structural states. The difference between apo and holo trajectories was particularly pronounced in variants with lower binding affinity: greater dynamical similarity between the bound and unbound states was associated with more stable binding.\\
Once binding occurred, residues at the S-ACE2 interface exhibited similar interaction patterns across most variants and the WT, as indicated by covariance analysis, irrespective of binding affinity. The only exception was the delta variant, which displayed a unique binding dynamics characterized by a reorientation around the ACE2 binding site.\\
Notably, binding affinity correlated with the variability of the intermolecular distance between the S protein and ACE2, with higher-affinity complexes exhibiting greater distance fluctuations.\\
Taken together, these observations suggest that a more stable binding interaction is achieved when the interaction follows a lock-and-key mechanism rather than a more dynamic conformational selection or induced-fit model. Variants with the highest binding affinity are more stable in isolation and exhibit similar dynamics between their apo and holo forms.
However, viral entry efficiency is not the sole determinant of SARS-CoV-2 fitness. Throughout the virus's evolution, certain variants exhibited more dynamical behavior, suggesting that other traits (such as antibody escape) were favored over maximal binding affinity.

\section*{Materials and Methods}

\subsection*{Structural data}

The structures of the isolated SARS-CoV-2 S protein, including both the WT and variants, as well as the complexes of the S protein bound to the human ACE2 receptor, were obtained from the Protein Data Bank (PDB). For our analysis, only the RBD of the S protein and the extracellular domain of ACE2 were considered. The corresponding PDB IDs are listed in Table \ref{tab_pdbid}.\\
We focused on the five variants for which the dissociation constant ($K_d$) of the S–ACE2 complex was experimentally determined by Han \textit{et al.} \cite{han2022receptor}, along with that of the WT. The corresponding values are reported in Table~\ref{tab_kd}.

\begin{table}[!ht]  
  \centering
  \begin{tabular}{lcc}
    \toprule
           & \textbf{APO Structure} & \textbf{HOLO Structure} \\ 
           \hline
    \textbf{Alpha} & 8DLI   & 8DLK \\
    \textbf{Beta} & 8DLL  &   8DLN  \\
    \textbf{Delta} & 7V7N   &  7W9I   \\
    \textbf{Gamma} & 8DLO  &  8DLQ  \\ 
    \textbf{Omicron} & 7Q07  & 7T9L  \\ 
    \textbf{WT} & 7DDD  & 6M0J  \\ 
    \hline
  \end{tabular}
  \caption{\textbf{Structural data.} For each of the considered variants and the WT, the first column reports the PDB ID of the isolated SARS-CoV-2 S protein, while the second column lists the PDB ID of the S in complex with ACE2.}
  \label{tab_pdbid}
\end{table}


\begin{table}[!ht]  
  \centering
  \begin{tabular}{lc}
    \toprule
           & \textbf{$K_d$ (nM)} \\ 
           \hline
               \textbf{WT} & 24.63  \\ 
    \textbf{Alpha} & 5.40 \\
    \textbf{Beta} &  13.83 \\
    \textbf{Delta} & 25.07  \\
    \textbf{Gamma} & 11.00  \\ 
    \textbf{Omicron} &  31.40 \\ 
    \hline
  \end{tabular}
  \caption{\textbf{Experimentally measured dissociation constants for each of the considered variant and the WT.} The $K_d$ of the S-ACE2 complex for each variant and the WT, as measured by Han \textit{et al.} \cite{han2022receptor}, is reported. Variants are reported in chronological order of appearance.}
  \label{tab_kd}
\end{table}




\subsection*{Molecular dynamics simulations}

All simulations were performed using Gromacs~\cite{gromacs}.
Topologies of the system were built using the CHARMM-36 force field~\cite{charmm}.
Each system was placed in a dodecahedric simulative box, with periodic boundary conditions, filled with TIP3P water molecules~\cite{Jorgensen1983}. For all simulated systems, we ensured that each atom of the proteins was at least $1.1 \, \mathrm{nm}$ from the box borders.
Each system was then minimized using the steepest descent algorithm. Next, a relaxation of water molecules and thermalization of the system were performed in NVT and NPT ensembles, each for $0.1 \, \mathrm{ns}$ at $2 \, \mathrm{fs}$ time-step. 
The temperature was maintained at $300 \, \mathrm{K}$ with v-rescale thermostat \cite{vrescale}; the final pressure was fixed at $1 \, \mathrm{bar}$ using the Parrinello-Rahman barostat \cite{parrinello}.
The LINCS algorithm \cite{lincs} was applied to constrain bonds involving hydrogen atoms.
A cut-off of $12 \, \angstrom$ was used for the evaluation of short-range non-bonded interactions, and the Particle Mesh Ewald method~\cite{Cheatham1995} was applied for the long-range electrostatic interactions. This procedure was followed for all the simulations.

\subsection*{Statistics and Reproducibility}
All molecular dynamics simulations were run for 500 ns, ensuring that each system reached equilibrium conformations. All subsequent analyses were performed by removing the first 50 ns of the simulations to focus on the equilibrium range.

\subsection*{Interface residue definition}

For each analyzed complex, the interface was defined by selecting ACE2 and SARS-CoV-2 S residues whose centroids are within 8 \AA~of each other in the average conformation. This average conformation is determined as the structure with the RMSD closest to the mean RMSD of the respective complex simulation.\\

For the analysis requiring a comparable set for all complexes, we performed two selections:
\begin{itemize}
    \item When comparing S protein holo and apo conformations, we considered all residues that appear in at least one of the interfaces, resulting in a total of 23 residues listed in Table \ref{tab_big_patch}. 
    This comprehensive selection allows for a complete characterization of the differences between the apo and holo conformations, as well as the variations among the different virus variants, including the interactions that are lost.
    \item To study the covariance between the S-ACE2 interfaces, we selected only the residues common to all interfaces, excluding those residue pairs that do not exhibit correlated motion due to structural changes in the binding sites. This resulted in 13 residues from ACE2 and 11 residues from the S protein, as listed in Table \ref{tab_interfaces}.
\end{itemize}

\begin{table}[!ht]  
  \centering
  \begin{tabular}{c}
    \toprule
          \textbf{S interacting residues} \\ 
           \hline
           LYS-417, TYR-453, LEU-455, PHE-456, TYR-473, GLN-474,\\
           ALA-475, GLY-476, SER-477, GLY-485, PHE-486, ASN-487,\\
           TYR-489, GLN-493, SER-494, GLY-496, GLN-498, THR-500, \\
           TYR-501, GLY-502, VAL-503, GLY-504, TYR-505 \\
    \hline
  \end{tabular}
  \caption{\textbf{spike residues identified as interacting in at least one of the variants or the WT.} Interacting residues are defined as those whose centroids are located within 8\AA~of ACE2 residues.}
    \label{tab_big_patch}
\end{table}

\begin{table}[!ht]
  \centering
  \begin{tabular}{p{6cm}|p{6cm}} 
    \toprule
    \textbf{S interface} & \textbf{ACE2 interface} \\ 
    \hline
    LEU-455, PHE-456, TYR-473, ALA-475, GLY-476, PHE-486, ASN-487, TYR-489, THR-500, GLY-502, VAL-503  
    & GLN-24, THR-27, ASP-30, LYS-31, HIS-34, TYR-41, LEU-79, MET-82, TYR-83, THR-324, LYS-353, GLY-354, ASP-355 \\ 
    \hline
  \end{tabular}
  \caption{\textbf{spike residues identified as interacting in all variants and WT.} Interacting residues are those whose centroids are within 8\AA~of ACE2 residues.}
  \label{tab_interfaces}
\end{table}

\subsection*{Frobenius distance}
The Frobenius norm of a matrix A is defined as
\begin{equation}
\|A\|_F = \sqrt{ \sum_{i=1}^{m} \sum_{j=1}^{n} |a_{ij}|^2 },\end{equation}
where $a_{ij}$ denotes the elements of the matrix.
If A represents the difference between two matrices X and Y, then $\|A\|_F$ provides a measure of the distance between X and Y.

\subsection*{Principal Component Analysis and Explained Variance}
PCA is a multivariate statistical technique employed to reduce the degrees of freedom in a dataset. This is achieved by transforming the original basis vectors describing the data into an orthogonal basis formed by the eigenvectors of the covariance matrix $\hat{C}$ associated with a set of observables. These eigenvectors, known as PCs, can be ranked in descending order based on their corresponding eigenvalues, which indicate the amount of variance each PC captures from the data.
By selecting the first $d$ PCs, which account for the highest variance, the dimensionality of the dataset can be effectively reduced while preserving most of its essential information.\\
A quantitative estimate of the information collected by each PC corresponding to an eigenvalue $\lambda_i$, is given by
the Explained Variance Ratio (EVR):
\begin{equation}\label{e evr}
    EVR(\lambda_i)=\frac{\lambda_i}{\sum_j^{N} \lambda_j},
\end{equation}
where $N$ is the number of original observables.

\section*{Data Availability}

The data that support the findings of this study are available from the corresponding author upon request.

\section*{Acknowledgements}
This research was partially funded by grants from ERC-2019-Synergy Grant (ASTRA, n. 855923); EIC-2022-PathfinderOpen (ivBM-4PAP, n. 101098989); Project `National Center for Gene Therapy and Drugs based on RNA Technology' (CN00000041) financed by NextGeneration EU PNRR
MUR—M4C2—Action 1.4—Call `Potenziamento strutture di ricerca e creazione di campioni nazionali di R\&S' (CUP J33C22001130001).

\bibliographystyle{unsrt}
\bibliography{mybibfile}

\clearpage
\appendix
\renewcommand{\thesection}{S\arabic{section}} 
\renewcommand{\thefigure}{S\arabic{figure}}   
\renewcommand{\thetable}{S\arabic{table}}     
\setcounter{section}{0}
\setcounter{figure}{0}
\setcounter{table}{0}

\section*{Supplementary Information}

\begin{figure*}[ht]
\centering
\includegraphics[width=\linewidth]{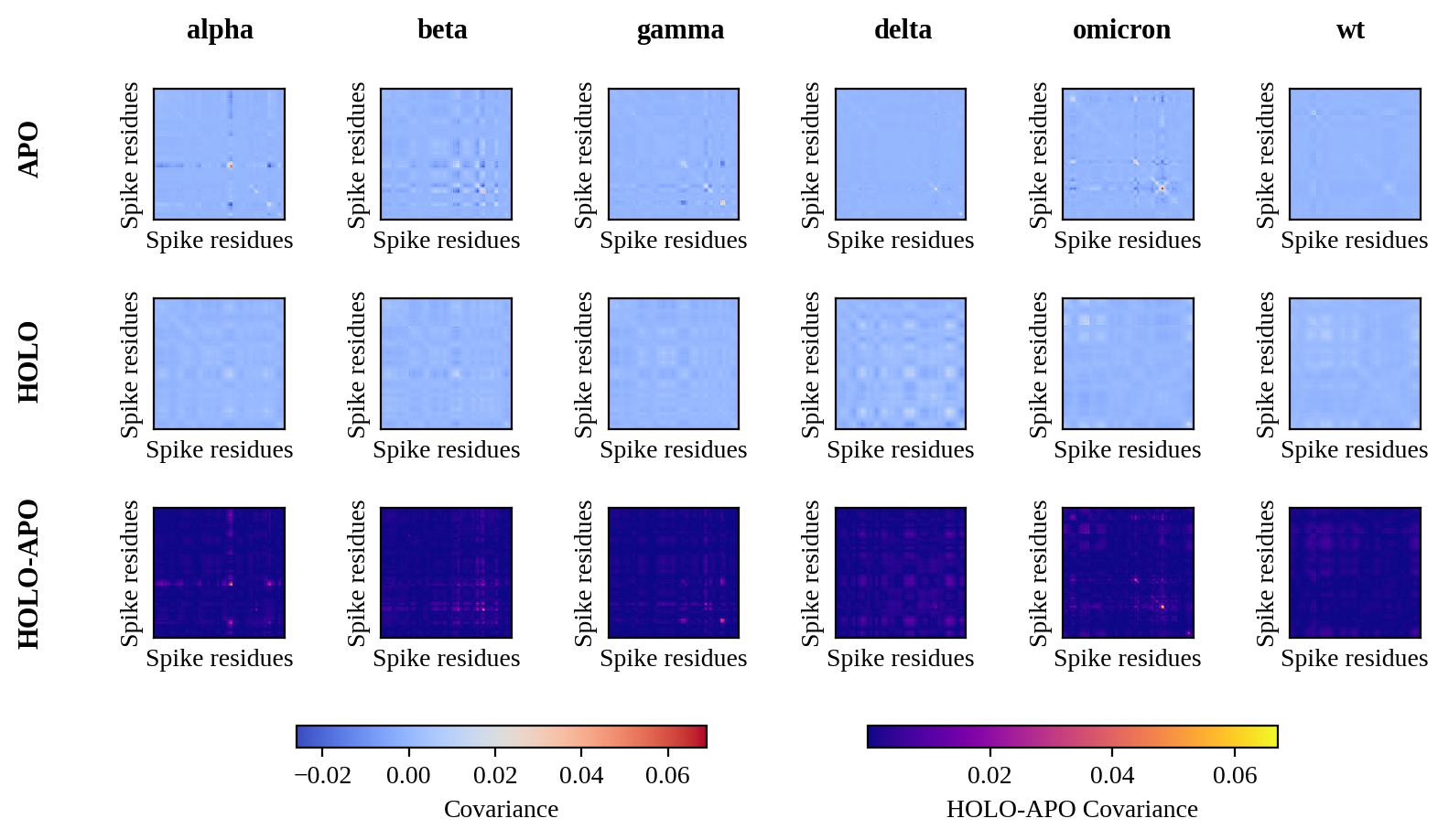}
\caption{\textbf{Covariant analysis of Spike residue motion in apo and holo conformations.}
For each variant and the wt, the covariance matrices of residue motions are presented across the first two rows. The top row displays the covariance values calculated from the apo conformation simulation, while the middle row represents those obtained from the holo conformation
The colormap represents the covariance values, as indicated by the left colorbar. The bottom row illustrates the difference in covariance between the holo and apo conformations for each residue pair. The right colorbar provides the corresponding scale for these differences.}
\label{figSI_1}
\end{figure*}

\begin{figure*}[ht]
\centering
\includegraphics[width=\linewidth]{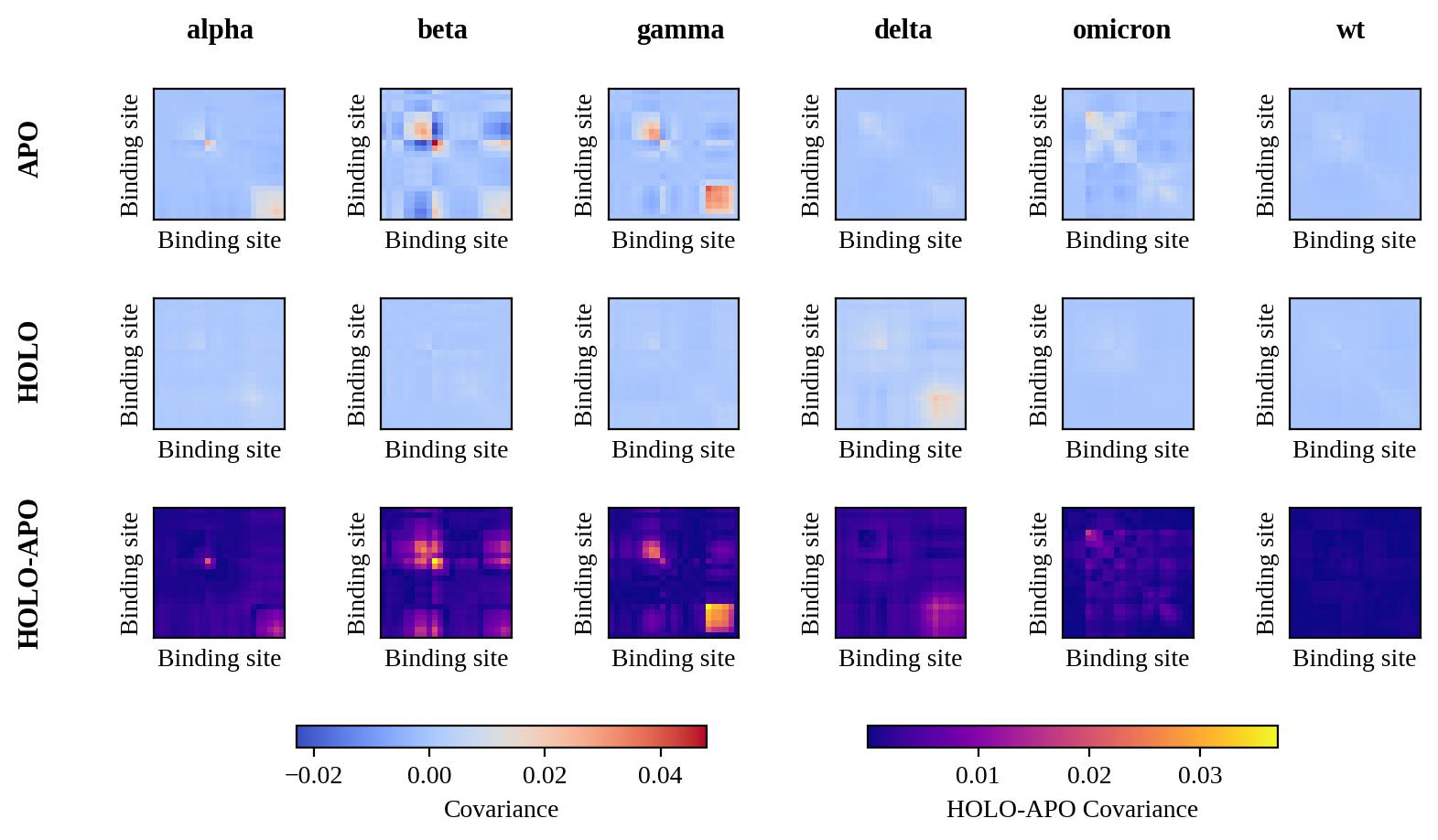}
\caption{\textbf{Covariant analysis of Spike binding residue motion in apo and holo conformations.}
The same analysis presented in Figure \ref{figSI_1} was performed considering only the binding site residues.
}
\label{figSI_2}
\end{figure*}

\begin{figure*}[ht]
\centering
\includegraphics[width=\linewidth]{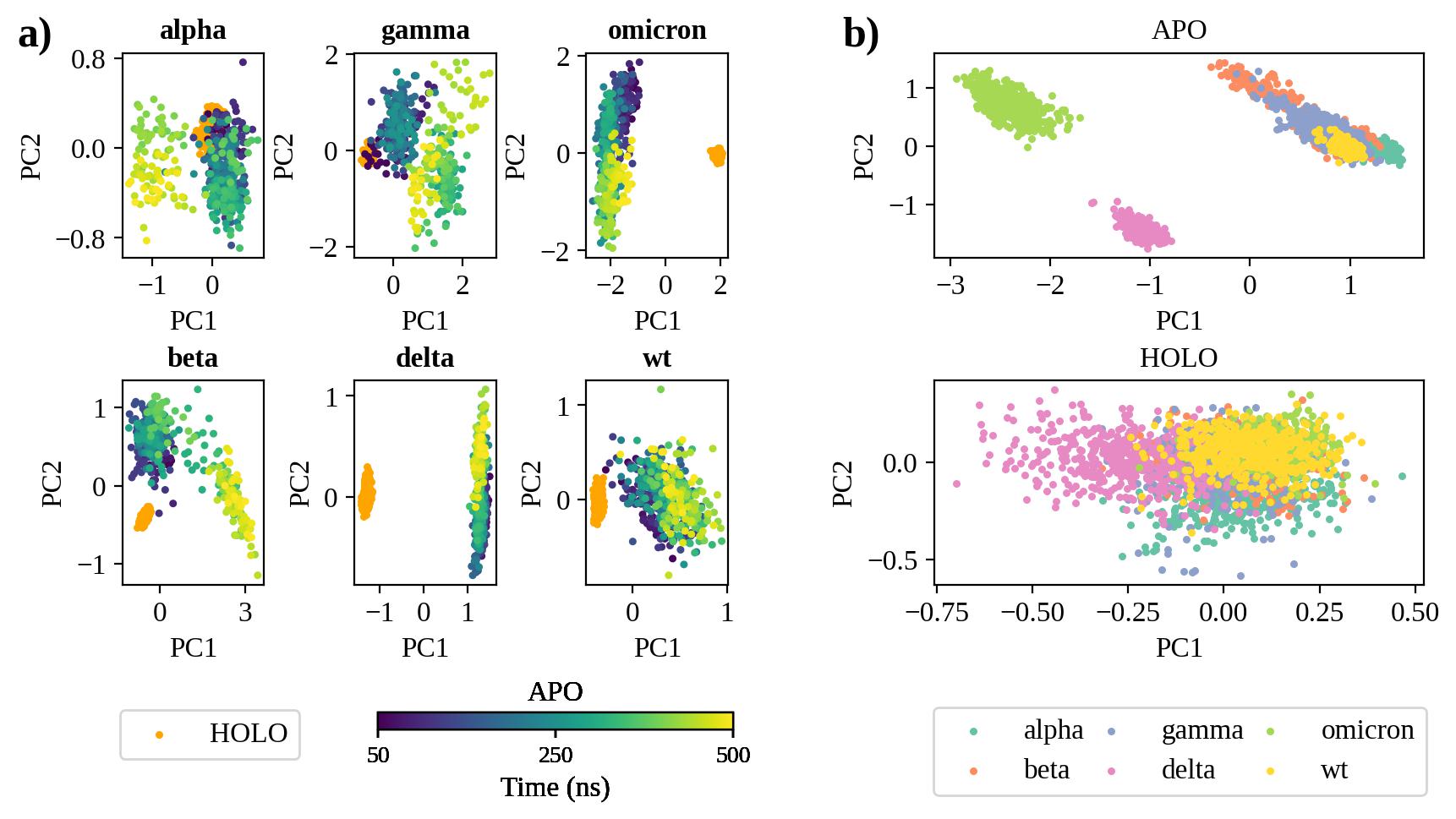}
\caption{\textbf{Principal Component Analysis of Spike binding residues in apo and holo trajectories.}
\textbf{a)} Two-dimensional projection of the sampled conformations in the subspace spanned by the first two PCs of atomic positions covariances during the apo and holo simulations, focusing on the residues of each variant and the wt interacting with ACE2. Points from holo simulations are shown in orange, while those from apo trajectories transition from blue to yellow as simulation time progresses from 0 to 500 ns, as indicated by the colorbar. The first 50 ns of simulation were excluded to account for the system’s initial non-equilibrium phase.
\textbf{b)} Two-dimensional projection of the sampled
conformations in the subspace spanned by the first two PCs of the atomic positions covariances, comparing apo (left) and holo
(right) simulation across all variants and wt, considering only the residues interacting with ACE2. Each point is colored according to its corresponding variant, as indicated in the legend.}
\label{figSI_3}
\end{figure*}

\begin{figure*}[ht]
\centering
\includegraphics[width=\linewidth]{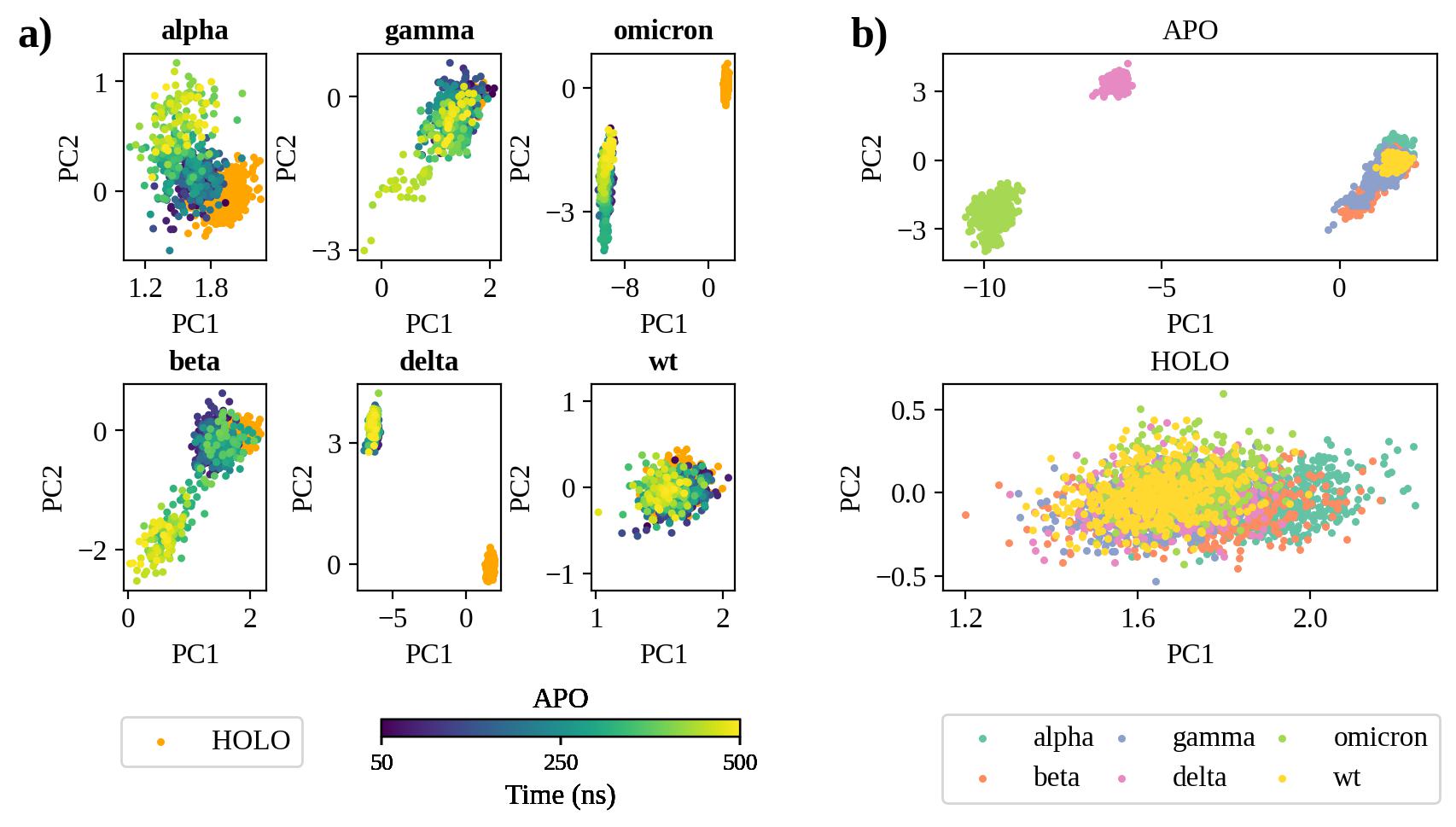}
\caption{\textbf{Full Principal Component Analysis of Spike residues in apo and holo trajectories.}
\textbf{a)} Two-dimensional projection of the sampled conformations in the subspace defined by the first two PCs of atomic positions covariances during the apo and holo simulation for each variant and the wt.
The PCA was performed on all the trajectories combined, so that the PCs are shared across all the plots.
Holo simulations points are colored in orange, while those
from apo trajectories are colored from blue to yellow as simulation time goes from zero to 500 ns, as indicated by the colorbar.
The first 50 ns of simulation were excluded from the analysis to account for the initial non-equilibrium phase of the system.
\textbf{b)} The same projection described in a) is here stratified by apo (left) and holo (right) simulations across all variants and the wt. Each point is colored according to its corresponding variant, as indicated in the legend.}
\label{figSI_4}
\end{figure*}

\begin{figure*}[ht]
\centering
\includegraphics[width=\linewidth]{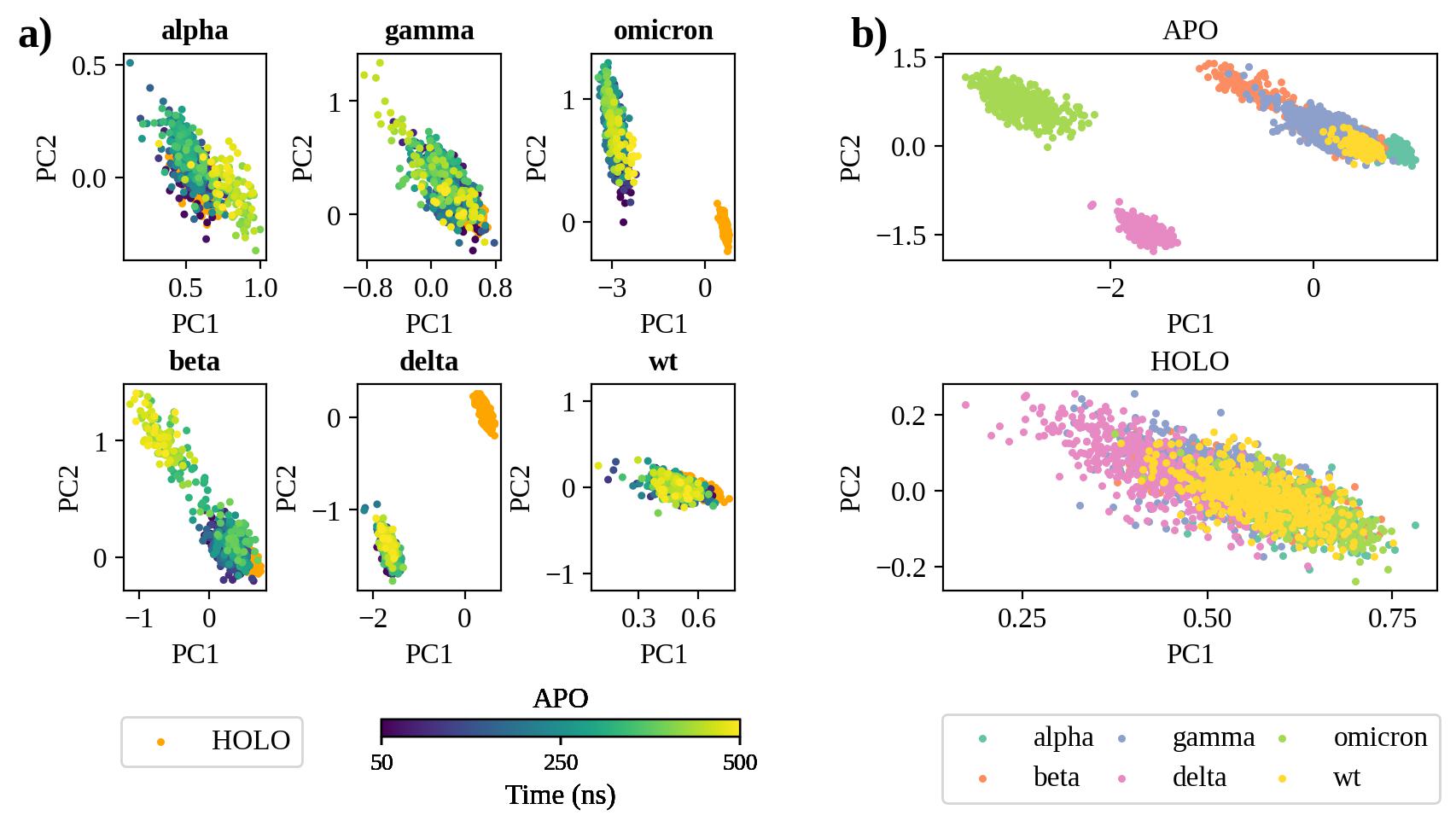}
\caption{\textbf{Full Principal Component Analysis of Spike binding residues in apo and holo trajectories.}
\textbf{a)} Two-dimensional projection of the sampled conformations in the subspace spanned by the first two PCs of atomic positions covariances during the apo and holo simulations for the residues of each variant and the wt interacting with ACE2.
The PCA was performed on all trajectories together, so that the PCs are shared across all the plots.
Holo simulations points are colored in orange, while those
from apo trajectories are colored from blue to yellow as simulation time goes from zero to 500 ns, as indicated by the colorbar.
The first 50 ns of simulation were excluded from the analysis to account for the initial non-equilibrium phase of the system.
\textbf{b)} The same projection as in a), now stratified in apo (left) and holo (right) simulations across all variants and wt. Each point is colored according to its corresponding variant, as indicated in the legend.}
\label{figSI_5}
\end{figure*}

\begin{figure*}[ht]
\centering
\includegraphics[width=\linewidth]{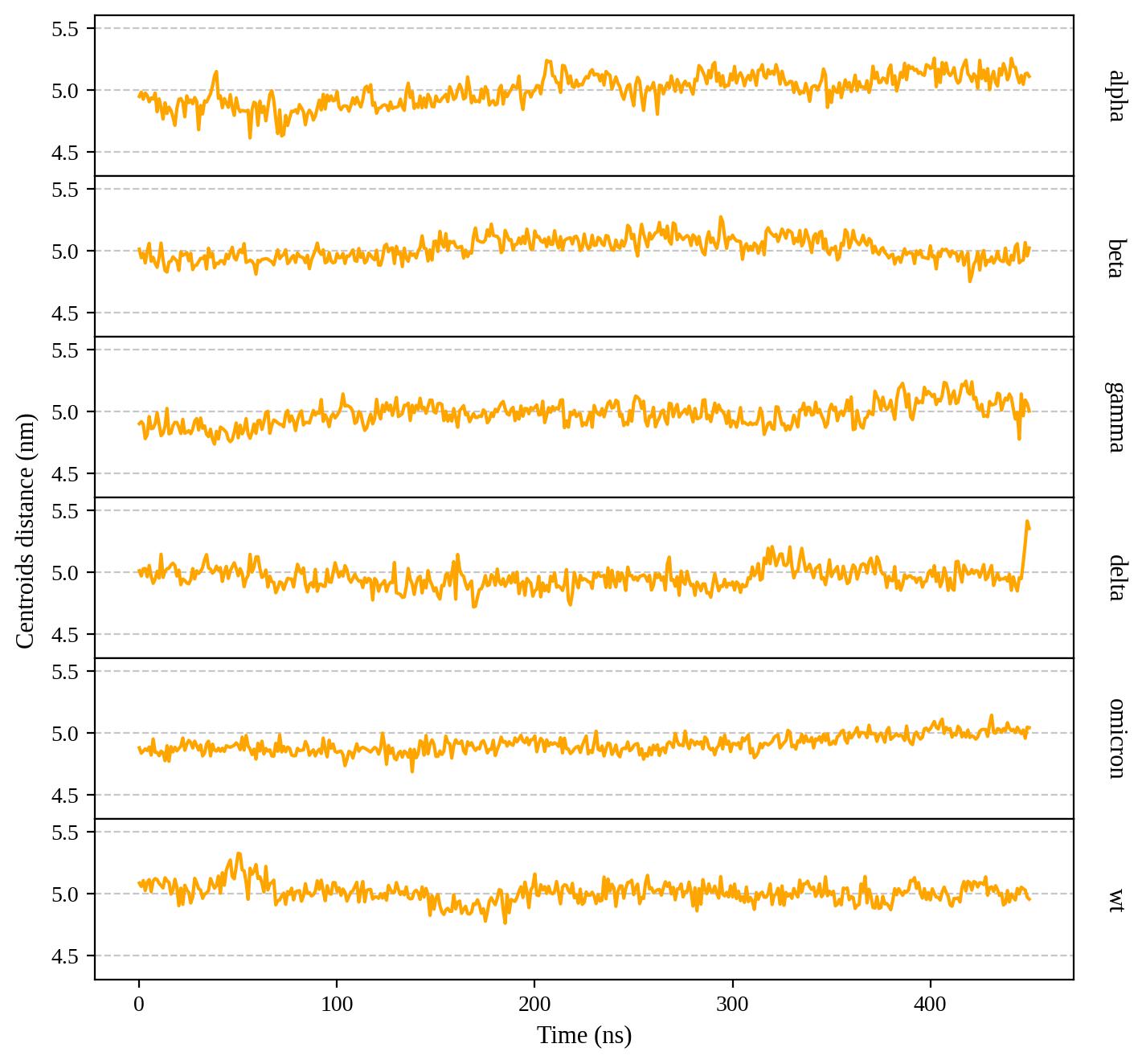}
\caption{\textbf{Relative motion of Spike and ACE2 in terms of centroids distance.}
For each variant and the wt, distance between the centroids of S protein and ACE2 is plotted as a function of the simulation time.}
\label{figSI_6}
\end{figure*}

\end{document}